\documentclass[twocolumn,showpacs,preprintnumbers,aps,prr,amsmath,amssymb,floatfix,superscriptaddress,longbibliography]{revtex4-2}

\usepackage{graphicx}
\usepackage{dcolumn}
\usepackage{bm}
\usepackage{epsfig}
\usepackage{color}
\usepackage{natbib}
\usepackage{physics}
\usepackage{amsmath}
\usepackage{hyperref}

\usepackage{ulem}

\def\Rb{$^{87}$Rb }
\def\Na{$^{23}$Na }

\def\NaRb{$^{23}$Na$^{87}$Rb }

\def\ket#1{\left|#1\right\rangle}
\def\Xstate{$X^1\Sigma^+$}

\begin{document}

\title{Observation of resonant dipolar collisions in ultracold $^{23}$Na$^{87}$Rb rotational mixtures}

\author{Junyu He}
\affiliation{Department of Physics, The Chinese University of Hong Kong, Hong Kong, China}
\author{Xin Ye}
\affiliation{Department of Physics, The Chinese University of Hong Kong, Hong Kong, China}
\author{Junyu Lin}
\affiliation{Department of Physics, The Chinese University of Hong Kong, Hong Kong, China}
\author{Mingyang Guo}
\thanks{Current address: 5. Physikalisches Institut and Center for Integrated Quantum Science and Technology, Universit\"{a}t Stuttgart, Pfaffenwaldring
57,70569 Stuttgart, Germany}
\affiliation{Department of Physics, The Chinese University of Hong Kong, Hong Kong, China}
\author{Goulven Qu{\'e}m{\'e}ner}
\email{goulven.quemener@universite-paris-saclay.fr}
\affiliation{Universit\'{e} Paris-Saclay, CNRS, Laboratoire Aim\'{e} Cotton, 91405, Orsay, France}
\author{Dajun Wang}
\email{djwang@cuhk.edu.hk} 
\affiliation{Department of Physics, The Chinese University of Hong Kong, Hong Kong, China}
\affiliation{The Chinese University of Hong Kong Shenzhen Research Institute, Shenzhen, China}

\date{\today}

\begin{abstract}

We report the investigation on dipolar collisions in rotational state mixtures of ultracold bosonic $^{23}$Na$^{87}$Rb molecules. The large resonant dipole-dipole interaction between molecules in rotational states of opposite parities brings about significant modifications to their collisions,
even when an electric field is not present. In this work, this effect is revealed by measuring the dramatically enhanced two-body loss rate constants in the mixtures. In addition, the dipolar interaction strength can be tuned by preparing the NaRb mixture in different rotational levels with microwave spectroscopy. When the rotational level combination is not of the lowest energy, contributions from hyperfine changing collisions are also observed. Our measured loss rate constants are in good agreement with a quantum close-coupling calculation which we also present in full detail.

\end{abstract}


\maketitle

\section{Introduction}

Following several decades of efforts, the research on ultracold polar molecules (UPMs) has become one of the most active areas in atomic, molecular and optical physics~\cite{bohn2017}. Breakthroughs in producing samples of ground-state UPMs have made available more and more molecular species~\cite{ni08,ta14,mo14,park15,guo16,rva17,se18,voges2020} which, with their different statistics, chemical reactivities and permanent electric dipole moments (PEDMs), are offering us a glimpse of their great potentials. As early as 2010, ultracold chemical reactions were observed for the first time in the pioneering $^{40}$K$^{87}$Rb experiment at JILA~\cite{ni10,ospelkaus2010}. Very recently, both the final and intermediate products following this reaction have been observed by the Harvard group~\cite{hu2019}, and revealed the selected parities of the final rotational states distribution in a magnetic field~\cite{hu2020}. For several UPMs without allowed chemical reaction channels, losses were still observed~\cite{ye18,gregory2019} and attributed to the formation and trapping light excitation of two-molecule complexes~\cite{gre20,liu2020}.            


Most of the resources of UPMs stem from their rich internal structures and PEDMs. The PEDM makes it possible for polar molecules to interact via the long-range and anisotropic dipole-dipole interaction (DDI). UPMs with strong DDIs have been proposed as a versatile platform for quantum simulation of strongly correlated many-body problems~\cite{yi2007,zoller2010} and robust quantum information processing~\cite{demille2002,ni18} which are not accessible with ultracold atoms interacting via short range forces.

As the PEDM exists only along the molecular axis, for polar molecules to interact via DDI, a non-zero effective dipole moment must first be induced in the laboratory frame. To this end, the most straightforward method is to place polar molecules inside a dc electric field which induces effective dipole moment by mixing between rotational levels. With both reactive KRb~\cite{ni10} and non-reactive NaRb UPMs~\cite{guo18} in moderate electric fields, strong effects of the DDI manifested as enhanced losses and have already been observed. However, to induce a large effective dipole moment, a rather high electric field with the exact value determined by the magnitude of the PEDM and the rotational constant is usually necessary. Incorporating such high electric fields into an ultracold system is feasible, as been done recently~\cite{matsuda2020}, but is still quite challenging as it requires an appropriate and advanced experimental setup~\cite{covey2018}. Another way is using an ac electric field to couple two rotational levels with opposite parities directly. For example, the two successive rotational levels $J=0$ and $J=1$ can be coupled by a microwave (MW) field, and strong DDI arises from the transition dipole moments (TDMs)~\cite{yan13,yan20}. 

\begin{figure*}[t]
\centering
\includegraphics[width=0.95 \linewidth]{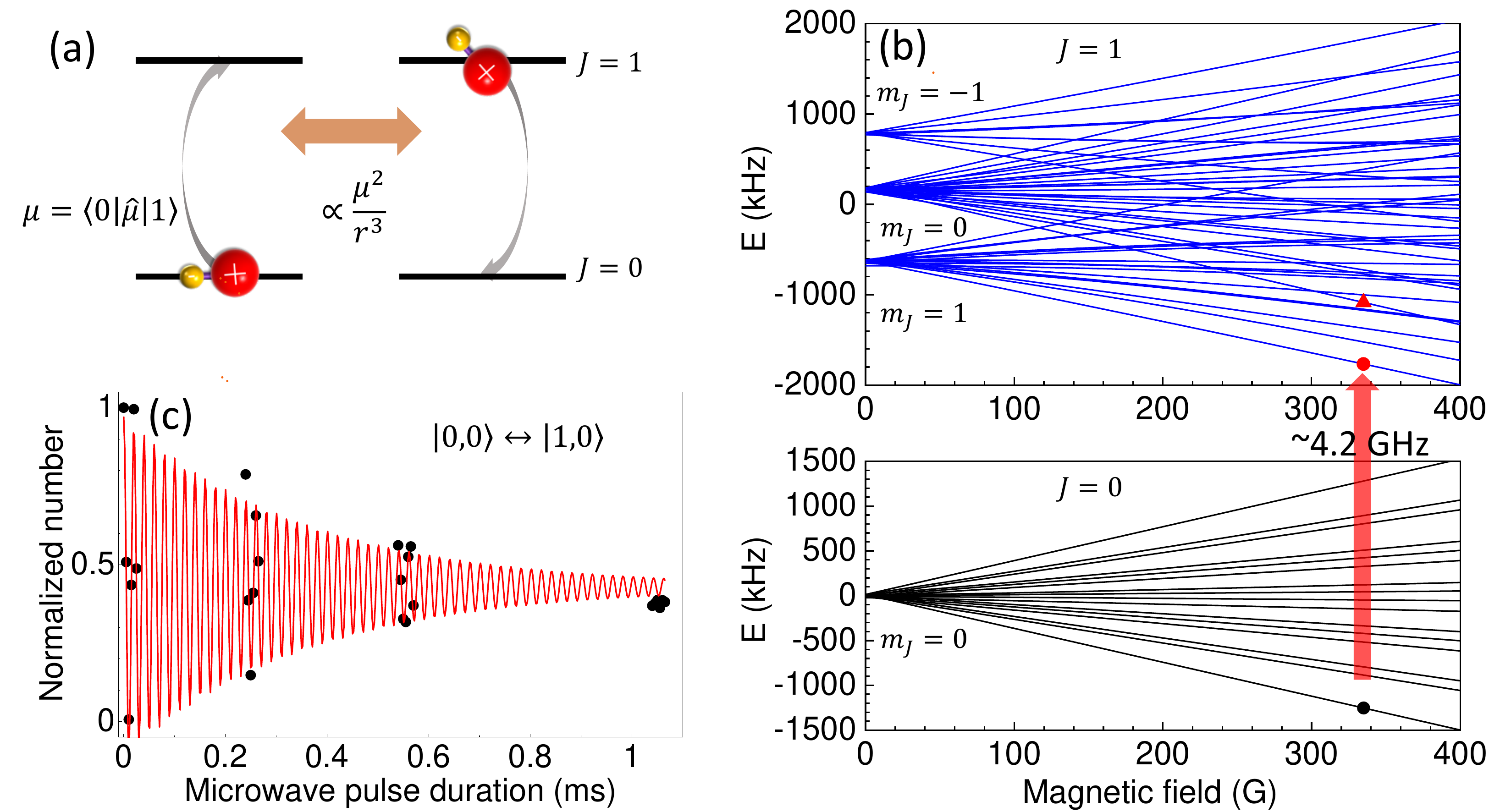}
\caption{Resonant dipole-dipole interaction between polar molecules in rotational levels of opposite parities. (a) The $J = 0$ and 1 rotational levels are connected by the TDM $\mu=\bra{0}\hat{\mu}\ket{1}$. Two molecules, with one prepared in each rotational level, can exchange the excitation resonantly which induces a strong DDI without any external electric fields. Here $\hat{\mu}$ is the dipole operator and $r$ is the inter-molecular distance. (b) Hyperfine structures of the $J = 0$ (lower panel) and $J = 1$ (upper panel) rotational levels of \NaRb molecule in magnetic field. Starting from molecules in the $\ket{0,0}$ level (black circle), the $\ket{1,1}$ (red triangle) or $\ket{1,0}$ (red circle) level can be populated selectively by applying a MW $\pi$-pulse and a 50-50 rotational mixtures can be prepared by a $\pi/2$-pulse (red arrow)~\cite{guo18a}. (c) Contrast decay of the MW driven Rabi oscillations for the $\ket{0,0} \leftrightarrow \ket{1,0}$ (bottom) transitions. The upper limits of the coherence time is 360~$\mu$s from the fittings (red curves). Similarly, the coherence time for the $\ket{0,0} \leftrightarrow \ket{1,1}$ transition is 590~$\mu$s. A coherent mixture prepared by a $\pi/2$-pulse will become incoherent after several milliseconds.}
\label{fig1}
\end{figure*}

In this work, we investigate strong dipolar collisions between ultracold \NaRb molecules in absence of external electric fields. To this end, we prepare the ultracold ground-state \NaRb molecules as an incoherent mixture of the $J=0$ and $J=1$ rotational levels. Due to the energy degeneracy of the dipole-dipole coupled pair states $(J =0, J = 1)$  and $(J = 1, J =0)$, two polar molecules with each prepared in these two rotational levels can interact directly via the so called $resonant$ dipole-dipole interaction [Fig.~\ref{fig1}(a)]. The resonant DDI has been well studied in ultracold Rydberg systems, for example by observing spin-exchange dynamics between single Rydberg atoms trapped in optical tweezers~\cite{sylain2017}. while the PEDM of UPMs is on the order of the Debye and is tens or even thousands times smaller than that of Rydberg atoms, UPMs have a much longer lifetime. It is thus still possible to implement high fidelity quantum gates with UPMs in optical tweezers~\cite{ni18,Sawant2020,Hughes2020}. 

In our investigation, the effect of the resonant DDI and its state dependence are revealed by measuring the loss rates of different rotational mixtures of \NaRb molecules. Compared with that between molecules prepared in a single internal state~\cite{ye18}, the loss in rotational mixtures is enhanced by about one order of magnitude. In addition, by comparing the measured loss rate with that from the quantum close-coupling modeling, contributions from hyperfine changing collisions are identified.

\section{Basic properties of ground-state \NaRb}
\label{hyperfine}

In the $X^1\Sigma^+$ ground electronic state, both the electronic orbital and spin angular momenta of the \NaRb molecule are zero. However, as both $^{23}$Na and $^{87}$Rb atoms have atomic nuclear spins $I = 3/2$, at the experimental magnetic field of 335 G, there are 16 and 64 hyperfine states in rotational level $J = 0$ and 1, respectively [Fig.~\ref{fig1}(b)]. Due to electronic quadrupole and nuclear spin-rotation couplings, these hyperfine states, labeled here as $\{\ket{J, m_J, m_I^{\rm Na}, m_I^{\rm Rb}}\}$, are often linear combination of the bare molecular states $\ket{J, m_J, m_I^{\rm Na}, m_I^{\rm Rb}}$. The curled brackets indicate that the molecular state is the eigenstate at a specific magnetic field and has a main character in the bare state $\ket{J, m_J, m_I^{\rm Na}, m_I^{\rm Rb}}$. Here $m_J$ is the projection of the rotational quantum number, and $m_I^{\rm Na}$ and $m_I^{\rm Rb}$ are the projections of the \Na and \Rb nuclear spins, respectively. The total angular momentum projection $m_F = m_J + m_I^{\rm Na} + m_I^{\rm Rb}$ is always a good quantum number and the coupling only happens between bare states with the same $m_F$. In this work, we use the three initial dressed states of increasing energies, given in term of the bare states with different amplitudes (admixtures)
\begin{eqnarray} 
\{{\ket{0,0,3/2,3/2}}\}  &=& - \ket{0,0,3/2,3/2}  \nonumber \\ 
\{{\ket{1,1,3/2,3/2}}\}  &=& \ket{1,1,3/2,3/2} \nonumber \\
\{{\ket{1,0,3/2,3/2}}\}  &=& - 0.906 \ket{1,0,3/2,3/2} \nonumber \\
 + \, &0.422& \,  \ket{1,1,3/2,1/2}  +  0.011 \ket{1,1,1/2,3/2} . \nonumber \\
\label{initstates}
\end{eqnarray}
The first two are the rotation and spin-stretch states with maximum possible $m_F$ for $J = 0$ and $J = 1$ and are identical to the corresponding bare states with $m_I^{\rm Na} = m_I^{\rm Rb} = 3/2$. The $\{{\ket{1,0,3/2,3/2}}\}$ state is a superposition state but with dominating $\ket{1,0,3/2,3/2}$ character. For simplicity, whenever possible without causing confusions, we will denote these three states as $\ket{0,0}$, $\ket{1,1}$, and $\ket{1,0}$, with respective $m_F=3$, 4, and 3. However, for the theoretical discussion, the full notation is still frequently used for clearness.

As shown in Fig.~\ref{fig1}(b), the $\ket{0,0}$ state with $J=0$ is the lowest energy level of the \NaRb molecule. For the two $J = 1$ states, $\ket{1,0}$ is the lowest energy level of this rotational manifold, while $\ket{1,1}$ lies above several other levels. Correspondingly, the $\ket{0,0} + \ket{1,0}$ two-body collision threshold has the lowest energy for $J = 0$ and $J = 1$ collisions. On the other hand, as the $\ket{0,0} + \ket{1,1}$ collision threshold is not the lowest one, inelastic relaxations to other hyperfine channels are possible.

In the electronic and vibrational ground state, the PEDM of NaRb molecules is $\mu_0 = 3.2$~D~\cite{guo16}. The effective resonant dipole moments reach $\mu_0/\sqrt{6} = 1.31$ D for the $\ket{0, 0}$ + $\ket{1, \pm 1}$ mixture, and $0.906\times\mu_0/\sqrt{3} = 1.67$~D for the $\ket{0,0}$ + $\ket{1,0}$ mixture. The additional 0.906 factor is a result of admixtures from other bare molecular states, see Eq.~\ref{initstates}. These effective dipole moments are much larger than all those have been reached in previous experiments~\cite{guo18}.

\section{Experiment and results}

\subsection{Preparation of the ground-state \NaRb sample}
Our experimental system has been discussed in detail before~\cite{ye18,guo18}. In brief, the experiment starts from ultracold samples of \NaRb molecules in the $\ket{0,0}$ rotational level of the electronic and vibrational ground state (\Xstate, $v = 0$) prepared by association of ultracold \Na and \Rb atoms to weakly bound Feshbach molecules followed by population transfer with a stimulated Raman adiabatic passage (STIRAP)~\cite{guo16}. The molecular sample is trapped in a cigar-shaped optical dipole trap formed by crossing two 1064.4~nm laser beams at a small angle of 27$^{\circ}$. The sample temperature can be tuned from 200~nK to about $2~\mu$K by varying the trap laser powers. For each trap power, we measure the trap oscillation frequencies from the sloshing motion of the ground-state \NaRb sample. In general, the $J = 1$ molecule should have different trap frequencies from those of the $J = 0$ molecule due to the anisotropic polarizability of the excited rotational levels. This effect is largely canceled as the polarizations of the two trap beams are orthogonal to each other. As a result, the measured trap frequencies are nearly identical for $J = 0$ and $J = 1$ samples.  

To prepare NaRb samples in different rotational states or their mixtures, we apply microwave pulses of different pulse areas~\cite{guo18a}. The $J = 0 \rightarrow J = 1$ transition frequency is around 4.2~GHz while the total spans of the hyperfine Zeeman structures of each rotational states are on the MHz level. For detection of $\ket{0,0}$ molecules, we first apply a reversed STIRAP to transfer the ground-state molecule back to Feshbach molecule. Following the magneto-dissociation, the number of atoms is detected with absorption imaging. For molecules in the $J = 1$ levels, we have to apply an additional MW $\pi$-pulse to transfer them back to the $\ket{0,0}$ level before the detection.

\subsection{Loss rate constants for molecules in single rotational states}
A complication for investigating collisions in UPM mixtures is the multiple loss channels. With the constant temperature assumption, the time evolutions of $N_0$ and $N_1$, the numbers of molecules in $J = 0$ and $J = 1$, are governed by the coupled rate equations 
\begin{align}
\frac{{\rm d}N_{0}(t)}{{\rm d}t} =-A \, \beta_{00} \frac{N_{0}(t)^{2}}{T^{3/2}}-A \, \beta_{01} \frac{N_{0}(t) N_{1}(t)}{T^{3/2}},\\
\frac{{\rm d}N_{1}(t)}{{\rm d}t} =-A \, \beta_{11} \frac{N_{1}(t)^{2}}{T^{3/2}}-A \, \beta_{01} \frac{N_{0}(t) N_{1}(t)}{T^{3/2}}.
\label{eq:mixture_twobody} 
\end{align}
Here, $\beta_{01}$, $\beta_{00}$, and $\beta_{11}$ are the loss rate constants between $J = 0$ and $J =1$, between pure $J = 0$ and between pure $J = 1$ molecules, respectively. $A=(\overline{\omega}^{2} m/4 \pi k_{B})^{3/2}$ is a constant with $k_B$ the Boltzmann constant, $m$ the mass of the molecule, and $\overline{\omega}$ the geometric mean of the trap frequencies. To measure $\beta_{01}$, contributions from $\beta_{00}$ and $\beta_{11}$ should be subtracted from the total loss. As the collisions for pure $J = 0$ and pure $J = 1$ samples are non-dipolar, the loss rate constants $\beta_{00}$ and $\beta_{11}$ are much smaller than $\beta_{01}$. In fact, it is this large difference which makes the $\beta_{01}$ measurement possible. In a separate experiment, we have tried to study the loss in the non-dipolar mixture of two hyperfine states of the $J = 0$ level. However, as the contributions from different channels are all comparable, we have not been able to extract the loss rate constant between the two hyperfine states reliably. 


In the current investigation, $\beta_{11}$ and $\beta_{00}$ are obtained in separate experiments. For measuring $\beta_{11}$, we start by preparing rotationally excited NaRb samples~\cite{guo18a} with a resonant $\pi$-pulse on the $\ket{0,0} \rightarrow \ket{1,0}$ or the $\ket{0,0} \rightarrow \ket{1,1}$ rotational transition [Fig.~\ref{fig1}(b)]. The transition frequencies are around 4.2 GHz and the frequency separation between $\ket{1,0}$ and $\ket{1,1}$ are 0.655 MHz~\cite{guo18a} in presence of a 335 G magnetic field. We then measure the molecule number versus holding time in the optical trap. We use only short holding time to avoid large temperature increase due to the heating effect. After fitting the data to the two-body loss model [Eq.~(2) without the $\beta_{01}$ term], we obtain $\beta_{11}$. We repeat the same measurement for sample temperatures from 250 nK to 1.5 $\mu$K to obtain the temperature dependence of the $\beta_{11}$ values for $\ket{1,0}$ and $\ket{1,1}$ samples. For $\beta_{00}$, the previously measured values are re-used~\cite{ye18}.

\begin{figure}[htbp]
\centering
\includegraphics[width=0.85 \linewidth]{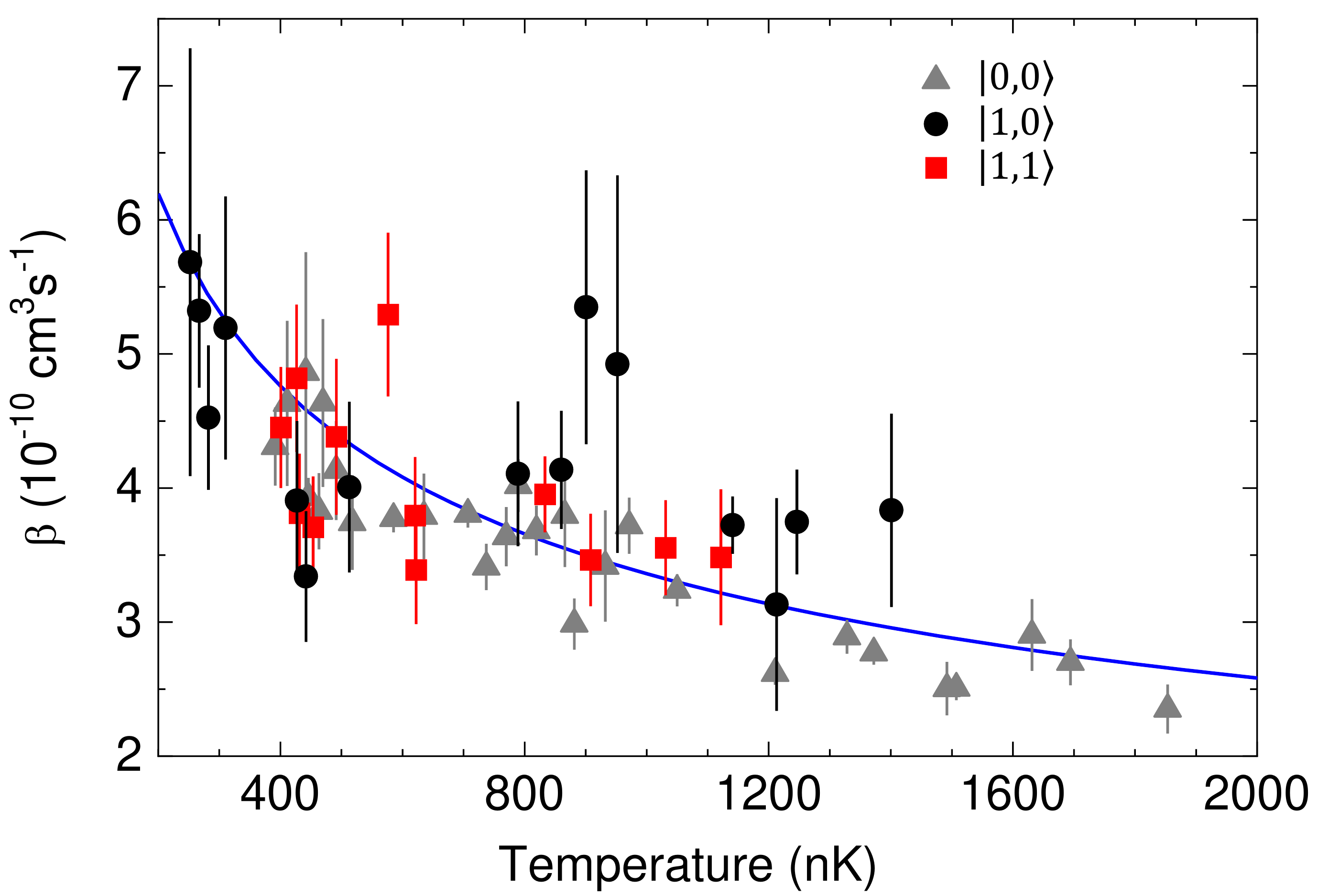}
\caption{Loss rate constants for collisions between \NaRb molecules in the same internal states. The $\beta_{11}$ values for the $\ket{1,0}$ (black dots) and $\ket{1,1}$ (red squares) samples are measured in this work, while $\beta_{00}$ are from~\cite{ye18}. The loss rate constants for the three cases show the same temperature dependence with similar values at the same temperature. For simplification purpose, they are represented with the same power law function of $T$ (blue curve) obtained in~\cite{ye18}. }
\label{fig2}
\end{figure}

Figure~\ref{fig2} summarizes the loss rate constants for \NaRb molecules in the three different levels which are very similar in both the magnitude and the temperature dependence. As have been discussed in~\cite{ye18,gregory2019}, these non-dipolar collisional losses are dominated by the complex formation process. The dynamics is governed by the van der Waals coefficient $C_6$ in the long range with a $C_6^{1/4}$ scaling~\cite{Idziaszek2010} and a typically large probability of complex formation in the short range. The $C_6^{1/4}$ scaling makes the long-range dynamics rather insensitive to possible $C_6$ variation with rotational levels. We do not know how the short-range complex formation probabilities change with rotational levels, but the results here hint that they are very similar. In what follows, we simply represent them with
\begin{equation}
\beta_{11} = \beta_{00} = \beta_0\times(T/T_0)^b,
\label{eq3}
\end{equation}
where $\beta_0=3.4\times10^{-10}$ cm$^3$ s$^{-1}$, $T_0$ = 0.97 $\mu$K and $b$ = $-0.38$~\cite{ye18}.

\subsection{Resonant dipole collisions in rotational mixtures}
To prepare the rotational mixtures, we apply a resonant ${\pi}/2$-pulse on the $\ket{0,0} \rightarrow \ket{1,0}$ or the $\ket{0,0} \rightarrow \ket{1,1}$ rotational transition. A short pulse duration of 6 $\mu$s is chosen so that losses during the pulse are negligible. To ensure the mixtures are incoherent, we wait for 3 ms before taking the loss data. As illustrated in Fig.~\ref{fig1}(c), the upper limits of the coherent time constants are less than 1 ms. An important issue here is the need for detecting molecules in both levels. Since each procedure takes several ms to complete, during the detection time for molecules in one level, molecules in the other level always experiences some additional losses. To avoid this complication, for each holding time we measure populations in the two rotational levels in two successive shots.

\begin{figure}[htbp]
\centering
\includegraphics[width=0.9 \linewidth]{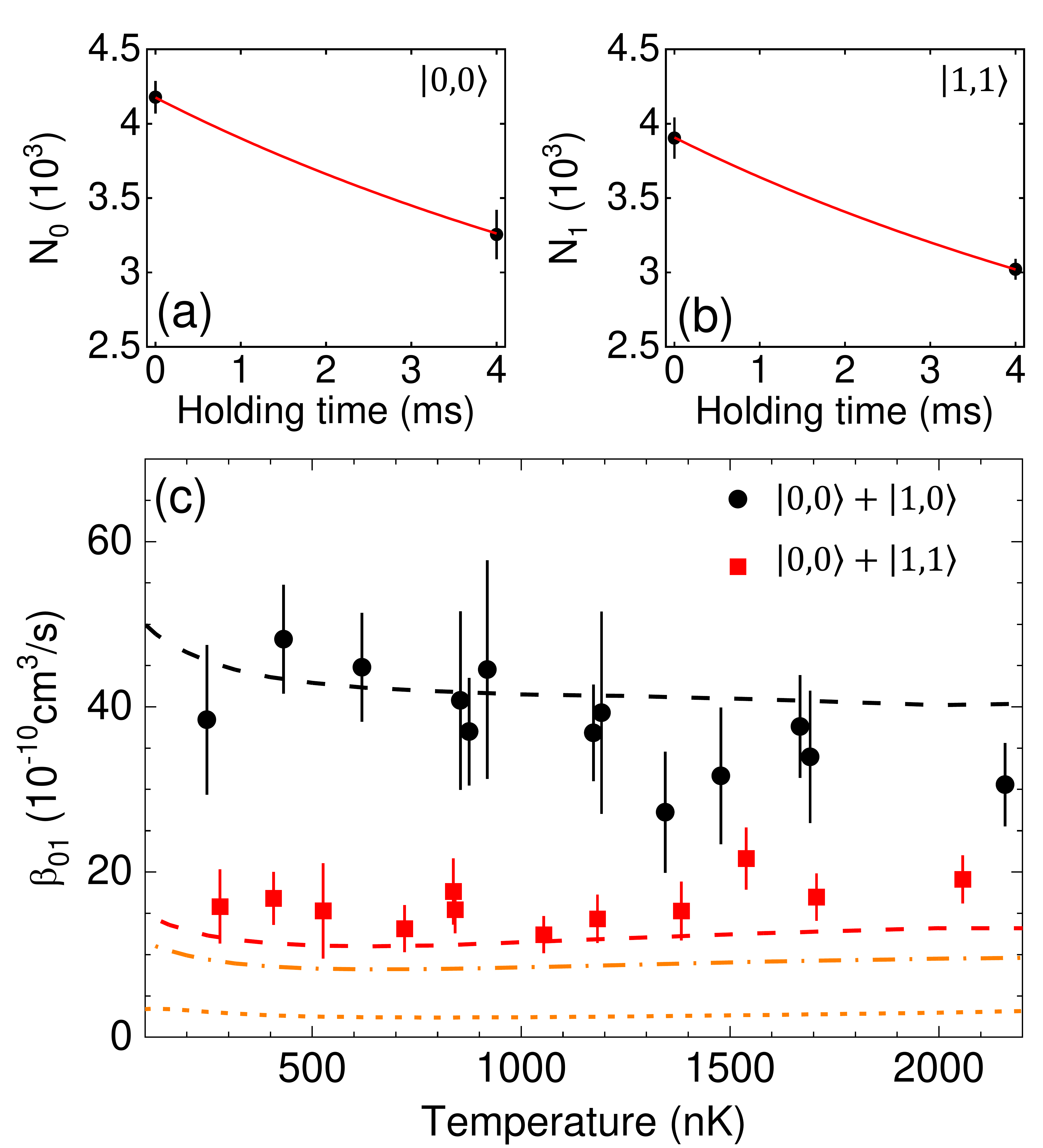}
\caption{Measurement of the loss rate constants between $J=0$ and $J=1$ molecules. The molecule numbers (a) $N_0$ and (b) $N_1$ are measured at two holding times. $\beta_{10}$ are extracted by fitting $N_0$ and $N_1$ simultaneously to Eq.~(1) and (2). (c) $\beta_{10}$ for $\ket{0,0} + \ket{1,0}$ (black dots) and $\ket{0,0} + \ket{1,1}$ (red squares). The dashed curves are the total loss rate constants for the two cases from close-coupling calculations. The orange dash-dot curve and the orange short-dash curve are contributions from complex formation and hyperfine changing for the $\ket{0,0}$ + $\ket{1,1}$ case, respectively. 
}
\label{fig3}
\end{figure}

As shown in Fig.~\ref{fig3}(a) and (b), we measure the number of molecules only at 0 ms and 4 ms. We use this short holding time to avoid the strong heating effect accompanying the rapid losses. As the loss rate constant is temperature dependent, heating in the course of the measurement always complicates the data analysis. Within 4 ms, the measured temperature change is less than 20\% and can be treated as a constant~\cite{ospelkaus2010,ye18,guo18}. To extract $\beta_{01}$, we fit these data points to Eqs.~(1) and (2) simultaneously with $\beta_{00}$ and $\beta_{11}$ fixed from the sample temperatures by Eq.~\eqref{eq3}. We have also verified experimentally that the $\beta_{01}$ values obtained from the four-point measurement match with those obtained from more points within 20\%.

Figure~\ref{fig3}(c) shows $\beta_{01}$ for collisions between two rotational mixtures, with black circles for the $\ket{0,0} + \ket{1,0}$, and red squares for the $\ket{0,0} + \ket{1,1}$. Comparing with the non-dipolar cases in Fig.~\ref{fig2}, the loss rate constants are 3 to 5 times higher for $\ket{0,0} + \ket{1,1}$ and 10 to 20 times higher for $\ket{0,0} + \ket{1,0}$. The ordering between the different magnitudes of the rates $\beta_{01}$ and the rate $\beta_{00}$ (or $\beta_{11}$) can be qualitatively explained by the DDI strengths achieved in this study, i.e., the higher the resonant dipole moment reached by the rotational mixture, the higher the loss rates, as more higher partial wave collisional channels are coupled in. The 1.67~D effective dipole moment for the $\ket{0,0}$ + $\ket{1,0}$ mixture is larger than the 1.31~D one for the $\ket{0,0}$ + $\ket{1,1}$ mixture, and of course larger than the non-dipolar cases $\beta_{00}$ or $\beta_{11}$, resulting in increasing losses in each cases.

\section{Theoretical modeling}

\subsection{The quantum close-coupling framework}

\paragraph{Nuclear spin states and symmetrized basis set}

For a quantitative understanding of the observations, we perform a time-independent quantum close-coupling calculation  \cite{Quemener_BookChapter_2018, Wang_NJP_17_035015_2015}, which has successfully described two-body collisions of pure $\ket{0,0}$ NaRb molecule \cite{ye18,guo18}. However, the rotational mixture collision is even more complicated to compute as both even and odd partial waves are allowed and the DDI is much larger. In addition, we need to include the proper hyperfine structure of the NaRb molecules at $B = 335$~G. As the nuclear spins couple with the rotation and interact with the magnetic field (see Sec.~\ref{hyperfine}), inelastic, hyperfine state changing transitions can occur, in addition to the loss caused by the complex formation. The experimentally observed loss should be the result of all these possible contributions. 
As described in Sec.~\ref{hyperfine}, 
the hyperfine dressed states at a given magnetic field $B$, are linear combinations of the bare molecular states~\cite{guo18a}
\begin{multline}\label{hypdressstate}
\{\ket{J, m_{J}, m_{I}^{\rm Na}, m_{I}^{\rm Rb}}\} = \sum \,
\ket{J, m_J, m_I^{\rm Na}, m_I^{\rm Rb}} \\
\times \langle J, m_J, m_I^{\rm Na}, m_I^{\rm Rb} |  \,  \{ | J, m_J, m_I^{\rm Na}, m_I^{\rm Rb} \rangle¨ \} .
\end{multline}
A list of different hyperfine dressed states of interest at $B=335$~G is given in the Appendix. 
For the collision between two molecules, 
one has to build combined molecular states (CMS), composed of two of these
hyperfine dressed states in Eq.~\eqref{hypdressstate}. They are denoted as
\begin{multline}\label{CMS}
\{|J_1, m_{J_1}, m_{I_1}^{\rm Na}, m_{I_1}^{\rm Rb} \rangle \, |J_2, m_{J_2}, m_{I_2}^{\rm Na}, m_{I_2}^{\rm Rb} \rangle \} = \\
\{|J_1, m_{J_1}, m_{I_1}^{\rm Na}, m_{I_1}^{\rm Rb} \rangle\} \, \{|J_2, m_{J_2}, m_{I_2}^{\rm Na}, m_{I_2}^{\rm Rb}  \rangle \} .
\end{multline}
When dealing with collisions of two identical molecules \cite{Quemener_BookChapter_2018}, 
one has to take care of
the interchange of the two identical molecules.
Eq.~\eqref{CMS} is recast into a symmetric and anti-symmetric CMS
characterized by $\eta = \pm 1$ respectively and noted
\begin{multline}\label{CMSsym}
\{|J_1, m_{J_1}, m_{I_1}^{\rm Na}, m_{I_1}^{\rm Rb} \rangle \, |J_2, m_{J_2}, m_{I_2}^{\rm Na}, m_{I_2}^{\rm Rb} \rangle \}_{\pm} = \\
\frac{1}{\sqrt{2}} \, \bigg( \{|J_1, m_{J_1}, m_{I_1}^{\rm Na}, m_{I_1}^{\rm Rb} \rangle\} \, \{|J_2, m_{J_2}, m_{I_2}^{\rm Na}, m_{I_2}^{\rm Rb}  \rangle \}  \\ 
\pm \{|J_2, m_{J_2}, m_{I_2}^{\rm Na}, m_{I_2}^{\rm Rb}  \rangle \} \, \{|J_1, m_{J_1}, m_{I_1}^{\rm Na}, m_{I_1}^{\rm Rb} \rangle\} \bigg).
\end{multline}
In addition, the partial waves $|l, m_l \rangle$ have also to be included to form the symmetrized collisional dressed basis set (also called scattering channels):
\begin{eqnarray} \label{coldresbasis}
\{|J_1, m_{J_1}, m_{I_1}^{\rm Na}, m_{I_1}^{\rm Rb} \rangle \, |J_2, m_{J_2}, m_{I_2}^{\rm Na}, m_{I_2}^{\rm Rb} \rangle \}_\pm \,  |l, m_l \rangle 
\end{eqnarray}
which from Eq.~\eqref{CMSsym} and Eq.~\eqref{hypdressstate} 
can be expressed in term of the bare molecular states.
Because the molecules are identical, 
the overall wavefunction follows the following selection rules. 
Permutation of two identical bosons leads to the selection rule $\eta \, (-1)^l = +1$.
This implies that symmetric CMS ($\eta =+1$) are associated with even partial waves $l$
and anti-symmetric CMS ($\eta =-1$) are associated with odd partial waves.
This is the case in this study for bosonic $^{23}$Na$^{87}$Rb molecules.
Conversely, permutation of two identical fermions leads to the selection rule $\eta \, (-1)^l = -1$.
$\eta =+1$ is associated with odd partial waves while $\eta =-1$ are associated with even partial waves.
\\

\paragraph{Dipole-dipole interaction}

The molecule-molecule collisions are driven by a long-range interaction described 
in the bare molecular states by the electric dipole-dipole interaction
\begin{widetext}
\begin{multline} \label{Vdd}
 \langle J_1, m_{J_1}, m_{I_1}^{\rm Na}, m_{I_1}^{\rm Rb} | \, \langle J_2, m_{J_2}, m_{I_2}^{\rm Na}, m_{I_2}^{\rm Rb} | \, \langle l, m_l | V_{{dd}}  | J_1', m_{J_1}', {m'}_{I_1}^{\rm Na}, {m'}_{I_1}^{\rm Rb} \rangle \, | J_2', m_{J_2}', {m'}_{I_2}^{\rm Na}, {m'}_{I_2}^{\rm Rb} \rangle \, | l', m_l' \rangle  =  \\
 \delta_{m_{I_1}^{\rm Na}, {m'}_{I_1}^{\rm Na}} \,  \delta_{m_{I_1}^{\rm Rb}, {m'}_{I_1}^{\rm Rb}}  \,  \delta_{m_{I_2}^{\rm Na}, {m'}_{I_2}^{\rm Na}}  \, \delta_{m_{I_2}^{\rm Rb}, {m'}_{I_2}^{\rm Rb}}  \\ 
 \times \sqrt{30} \, \frac{\mu_0^2}{4 \pi \varepsilon_0 \, r^3} \,
  \sum_{p_1=-1}^{1} \sum_{p_2=-1}^{1} \sum_{p=-2}^{2}
   (-1)^{m_{J_1}+m_{J_2}+m_l}
 \, \left( \begin{array}{ccc}  1 & 1 & 2 \\ p_1 & p_2 & -p \end{array}  \right) \\
\times  \sqrt{(2J_1+1) \, (2J_1'+1)}
\, \left( \begin{array}{ccc} J_1 & 1 & J_1' \\ 0 & 0 & 0 \end{array} \right)
\, \left( \begin{array}{ccc} J_1 & 1 & J_1' \\ -m_{J_1} & p_1 & m_{J_1}' \end{array}  \right)  \\
\times  \sqrt{(2J_2+1) \, (2J_2'+1)}
\, \left( \begin{array}{ccc} J_2 & 1 & J_2' \\ 0 & 0 & 0 \end{array} \right)
\, \left( \begin{array}{ccc} J_2 & 1 & J_2' \\ -m_{J_2} & p_2 & m_{J_2}' \end{array}  \right) \\
 \times  \sqrt{(2l+1) \, (2l'+1)}
\, \left( \begin{array}{ccc} l & 2 & l' \\ 0 & 0 & 0 \end{array} \right)
\, \left( \begin{array}{ccc} l & 2 & l' \\ -m_{l} & -p & m_{l}' \end{array}  \right).
\end{multline}
\end{widetext}
Only the rotational and partial waves quantum numbers are involved in this expression, as the inter-atomic distance between the two different atoms in the heteronuclear molecule carries the permanent dipole moment $\mu_0$ and the dipole-dipole interaction depends on the mutual directions of the two molecules. 
The hyperfine quantum numbers are not involved and simply acts as spectators in Eq.~\eqref{Vdd}. But this is expressed in the bare molecular states basis set. In our formalism, we use the symmetrized collisional dressed basis set Eq.\eqref{coldresbasis}. With the help of Eq.~\eqref{CMSsym} and because the molecular dressed states have different bare states characters (Eq.~\eqref{hypdressstate}), couplings between different combined molecular dressed states via the dipole-dipole interaction are possible, responsible for inelastic, hyperfine state changing transitions in collisions. \\

\paragraph{Symmetries and good quantum numbers}

For each rotational state $J$, there are $(2I + 1)^2(2J + 1)$ hyperfine eigenstates.
When two colliding molecules are considered (with different values $J_1\ne J_2$), 
the number of scattering channels 
for a given partial wave $l$ amounts 
$(2I +1)^2(2J_1 + 1)(2I +1)^2(2J_2 + 1) / 2 \times (2l+1)$ for each 
symmetrized states $\eta=\pm1$.
The good quantum numbers are: 

(i) $M = m_{F_1} + m_{F_2} + m_l$, 
where $m_{F_1} = m_{J_1} + m_{I_1}^{\rm Na} + m_{I_1}^{\rm Rb}$ 
and $m_{F_2} = m_{J_2} + m_{I_2}^{\rm Na} + m_{I_2}^{\rm Rb}$ are the projection of the individual molecules 1 and 2 onto the quantization axis,

(ii) the parity $\epsilon_i=(-1)^{J_1+J_2+l}=\pm 1$
as no electric field is present, 

(iii) $\eta=\pm1$ characterizing a symmetric/anti-symmetric state of two combined molecules, as the study deals with identical molecules. 
\\

\noindent To account properly for the rotational $C_6^{rot}$ coefficient, one has to include at least a rotational basis set with quantum numbers $J=0,1,2$ and the corresponding values of $m_J$ where $-J \le m_J \le +J$. One cannot use a unique, effective rotational $C_6^{rot}$ because 
there will be as much values as there are combined molecules states. 
As a consequence, 
there are a lot of CMS to treat in the collisional problem. One also has to include the partial waves $|l, m_l \rangle$. The typical values used here are $l = 0,~2,...10$ for the $(\epsilon_i=-1, \eta = +1)$ symmetry and $l = 1,~3,...11$ for the $(\epsilon_i = +1, \eta = -1)$ symmetry, with $-l \le m_l \le +l$, to converge the results at the temperatures covered by the experiment.
For the states of interest involved in the experimental mixture collision, one has to consider 
the rotational state couples $(J_1,J_2) = (0,1)$ and $(1,2)$. They correspond exclusively to the symmetries 
($\epsilon_i=-1, \eta=+1$) and ($\epsilon_i=+1, \eta=-1$).
For the symmetry ($\epsilon_i=-1, \eta=+1$), the couple (0,1) corresponds to 
$(2I +1)^2(2J_1 + 1) (2I +1)^2(2J_2 + 1) / 2 \times \sum_{l} (2l+1) = 384 \times 66= 25344$ channels
and the couple (1,2) corresponds to 
$1920 \times 66= 126720$ channels. 
Similarly, for the symmetry ($\epsilon_i=+1, \eta=-1$), the couple (0,1) corresponds to 
$384 \times 78 = 29952$ channels
and the couple (1,2) corresponds to 
$1920 \times 78 = 149760$ channels. 
This results in $\sim 10^5$ 
scattering channels for the different symmetries,
providing matrices with $\sim 10^{10}$ elements.
Despite the existence of the previous good quantum numbers, 
this becomes easily an intractable numerical problem 
and a theoretical support for the experimental observations would not be possible. 
To overcome this problem, we proceed to a relevant simplification of the problem.  
\\

\paragraph{Selection of the relevant nuclear spin states}

A clever selection of the nuclear spin states is then performed. 
It consists in including dressed states in the basis set Eq.~\eqref{coldresbasis} 
that contain significant admixture of the bare states that compose the initial dressed state given
in Eq.~\eqref{initstates}.
More precisely, we keep exclusively the dressed states 

(i) which linear combination contain at least a bare state 
with $m_{I}^{\rm Na} = 3/2, m_{I}^{\rm Rb} =1/2$ 
and/or $m_{I}^{\rm Na} = 3/2, m_{I}^{\rm Rb} =3/2$ ;

(ii) with a character amplitude (in absolute value) of such bare states higher than a cut-off value, 
chosen in this study to be 0.38. 
\\

\noindent All other dressed states are removed from the problem.
This enables to reduce considerably the number of channels 
by removing many unnecessary ones while preserving
the main relevant admixtures of the scattering channels 
responsible for inelastic, hyperfine state changing transitions. 
We made this choice because the initial states $\{| 1, 1, 3/2, 3/2\rangle\}$ and $\{| 1, 0, 3/2, 3/2\rangle\}$ have essentially main character amplitudes 
(higher than the cut-off value taken as 0.38 here) of the bare states such that 
$m_{I}^{\rm Na} = 3/2, m_{I}^{\rm Rb}=1/2$ and/or $m_{I}^{\rm Na}= 3/2, m_{I}^{\rm Rb}=3/2$,
see Eq.~\eqref{initstates}. In this way, we preserve the character of the initial colliding states.
In doing so, we remove unnecessary, uncoupled or barely coupled states. For example the state 
$\{| 1, 0, 1/2, 3/2\rangle\}$ (see list in the Appendix) 
is removed as it has main character amplitudes of the bare states $m_{i_{\rm{Na}}} = 1/2, m_{i_{\rm{Rb}}}=3/2$ and $m_{i_{\rm{Na}}} = 1/2, m_{i_{\rm{Rb}}}=1/2$ and a low character amplitude (smaller than the cut-off 0.38) of the bare states 
$m_{i_{\rm{Na}}} = 3/2, m_{i_{\rm{Rb}}}=1/2$ and $m_{i_{\rm{Na}}} = 3/2, m_{i_{\rm{Rb}}}=3/2$.
It is expected that couplings via the dipole-dipole interaction in Eq.~\eqref{Vdd}
between the dressed states $\{| 1, 0, 1/2, 3/2\rangle\}$ and the intial ones 
$\{| 1, 1, 3/2, 3/2\rangle\}$ or $\{| 1, 0, 3/2, 3/2\rangle\}$ to be small due to the small hyperfine character overlap and conversely due to the high mismatch, so that we can neglect them. 
This argument can be seen in Fig.~\ref{fig8} and Fig.~\ref{fig9} in the Appendix where we discuss
the validity of this specific selection. The removal of the unnecessary channels
barely changes the adiabatic energies (which act as effective potentials for the dynamics),
at least at ultracold collision energies.
Therefore, the cross sections and rate constants will remain mainly the same. This is a good approximation one has to make to compute numerically feasible theoretical results for the experimental observations. \\

\paragraph{Solving the Schr{\"o}dinger equation.}

An usual time-independent partial-wave expansion of the colliding wavefunction is employed leading to a set of differential coupled Sch{\"o}dinger equations  \cite{Quemener_BookChapter_2018,Wang_NJP_17_035015_2015,ye18,guo18a} which is solved using the method of the log-derivative propagation, from an inter-molecular distance $r_{min}=5~a_0$ to $r_{max}=100000~a_0$, where $a_0$ is the Bohr radius.
Matching with asymptotic boundary conditions, the scattering matrix is obtained and the cross sections and rate constants for a given initial state of the molecules is extracted for a given collision energy and magnetic field. 
To find the temperature dependence of the rate constants, as plotted in Fig.~\ref{fig3}(c), we average the cross sections over a Maxwell--Boltzmann distribution of the relative velocities. 
To account for the complex formation losses of the NaRb molecules, a short-range boundary condition is introduced so that when the molecules meet in this region, they are lost with a unit probability. We use this condition because the large effective dipole moments found in the two rotational mixtures are equivalent to applying large electric fields, where the unit probability condition at short-range has shown to be consistent with previous experimental data~\cite{guo18}. 

\subsection{Total rate constants}

\begin{figure}[t]
\centering
\includegraphics[width = 0.9 \linewidth]{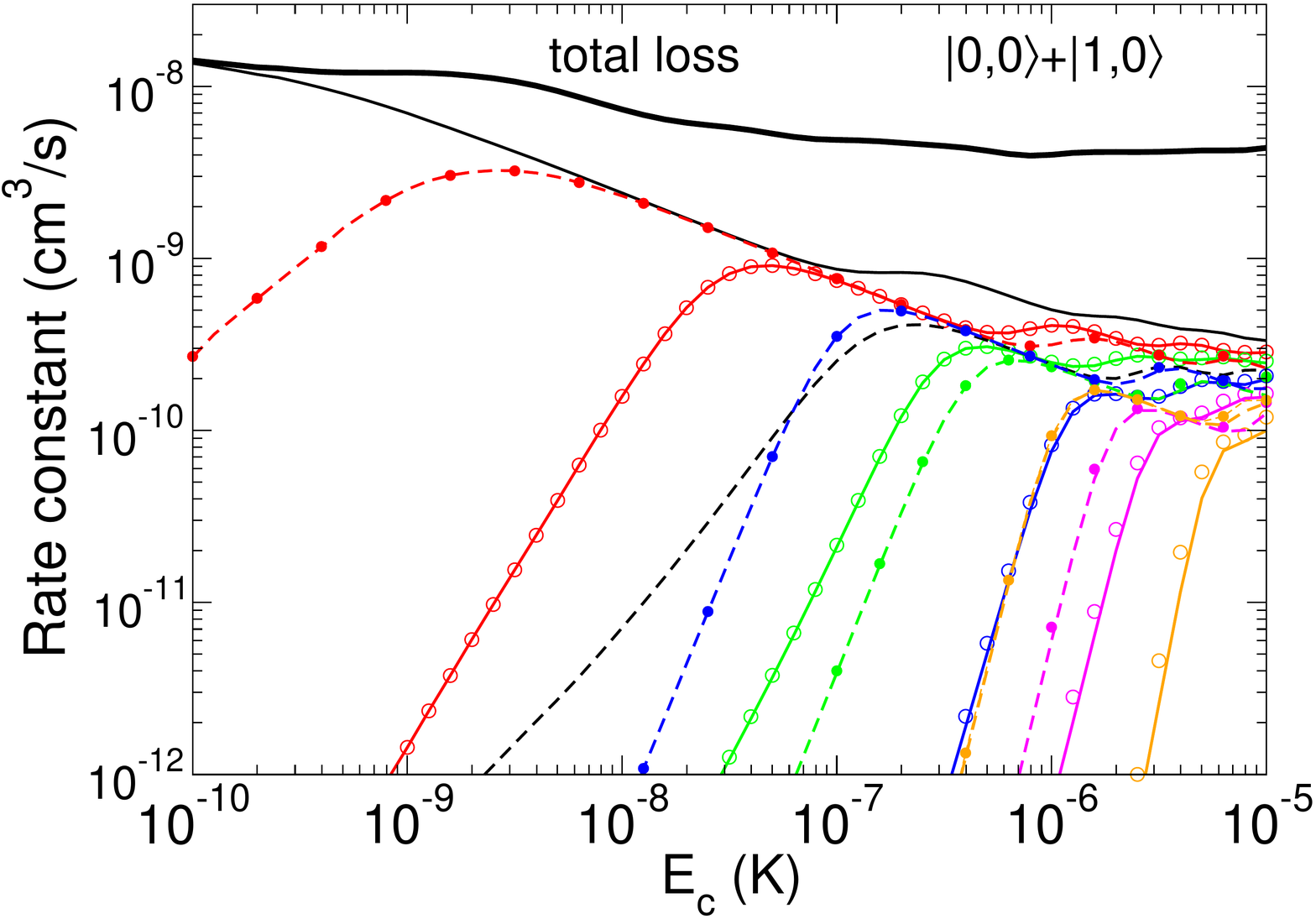} 
\caption{Loss rate constants $\beta_{01}$ for the mixture
$\ket{0,0}$ + $\ket{1,0}$ as a function of the collision energy, from close-coupling calculations. Thin solid curves and filled circles are for the complex formation loss rate constants with the symmetry $(\epsilon_i=-1, \eta=+1)$ corresponding to $l = 0,~2,... 10$. Thin dashed curves and open circles are for those with $(\epsilon_i=+1, \eta=-1)$ corresponding to $l = 1,~3,... 11$. 
Contributions from partial waves $m_l =-5,-4,...+5$ are included for both symmetries. 
The color code is the following: black curve ($m_l=0$), 
red curves ($m_l=+1$), red circles ($m_l=-1$),
green curves ($m_l=+2$), green circles ($m_l=-2$),
blue curves ($m_l=+3$), blue circles ($m_l=-3$),
magenta curves ($m_l=+4$), magenta circles ($m_l=-4$),
orange curves ($m_l=+5$), orange circles ($m_l=-5$).
The black thick solid line is the total loss rate constant and corresponds to the sum of all these 22 curves.}
\label{fig4}
\end{figure} 

Using the previous selection of the hyperfine dressed states, 
the values of $J=0,1,2$ and the partial waves $l = 0,~2,...10$ or $l = 1,~3,...11$,
we compute the rate constants 
at a magnetic field of $B = 335$~G. 
%
%
Due to the symmetry constraints, the $J=0$ + $J=1$ types of pair cannot relax to the 
$J=0$ + $J=0$ ones via collisions, as they belong to different $(\epsilon_i, \eta)$ symmetries.
Indeed, for $\eta=+1$ (even $l$ for identical bosons), 
the types of pair $J=0$ + $J=1$ correspond to $\epsilon_i = -1$
while the ones $J=0$ + $J=0$ correspond to $\epsilon_i = +1$.
Conversely, for $\eta=-1$ (odd $l$ for identical bosons), 
the types of pair $J=0$ + $J=1$ correspond to $\epsilon_i = +1$
while the ones $J=0$ + $J=0$ correspond to $\epsilon_i = -1$.
In addition, $\ket{0,0}$ + $\ket{1,0}$ with $m_{I}^{\rm Na} = m_{I}^{\rm Rb} = 3/2$ is the lowest collisional channel of the $J=0$ + $J=1$ threshold. Thus, there are no other loss processes but complex formation for this mixture. 
For each $\epsilon_i=\pm1$ parity, we compute the 11 curves corresponding to the rate constant 
for $m_l=-5,-4,..,+4,+5$ as a function of the collision energy. 
For the $\ket{0,0}$ + $\ket{1,0}$ mixture, these $m_l$ values 
correspond to $M=1,2,...,10,11$ as $m_{F_1} = m_{F_2} = 3$.
The $m_l=0$ contribution corresponds to $M=6$. 
This results in 22 curves in total which are added together to obtain the complex formation loss rate constant corresponding to the total loss rate constant.
This is reported in Fig.~\ref{fig4} as a function of the collision energy
and in Fig.~\ref{fig3}(c) as a function of the temperature.
One can see that due to the strong dipole-dipole interaction, 
the $s$-wave regime, where the rate constants
tend to a constant value at vanishing collision energies, occurs at 0.1~nK or below. \\

\begin{figure}[t]
\centering
\includegraphics[width = 0.9 \linewidth]{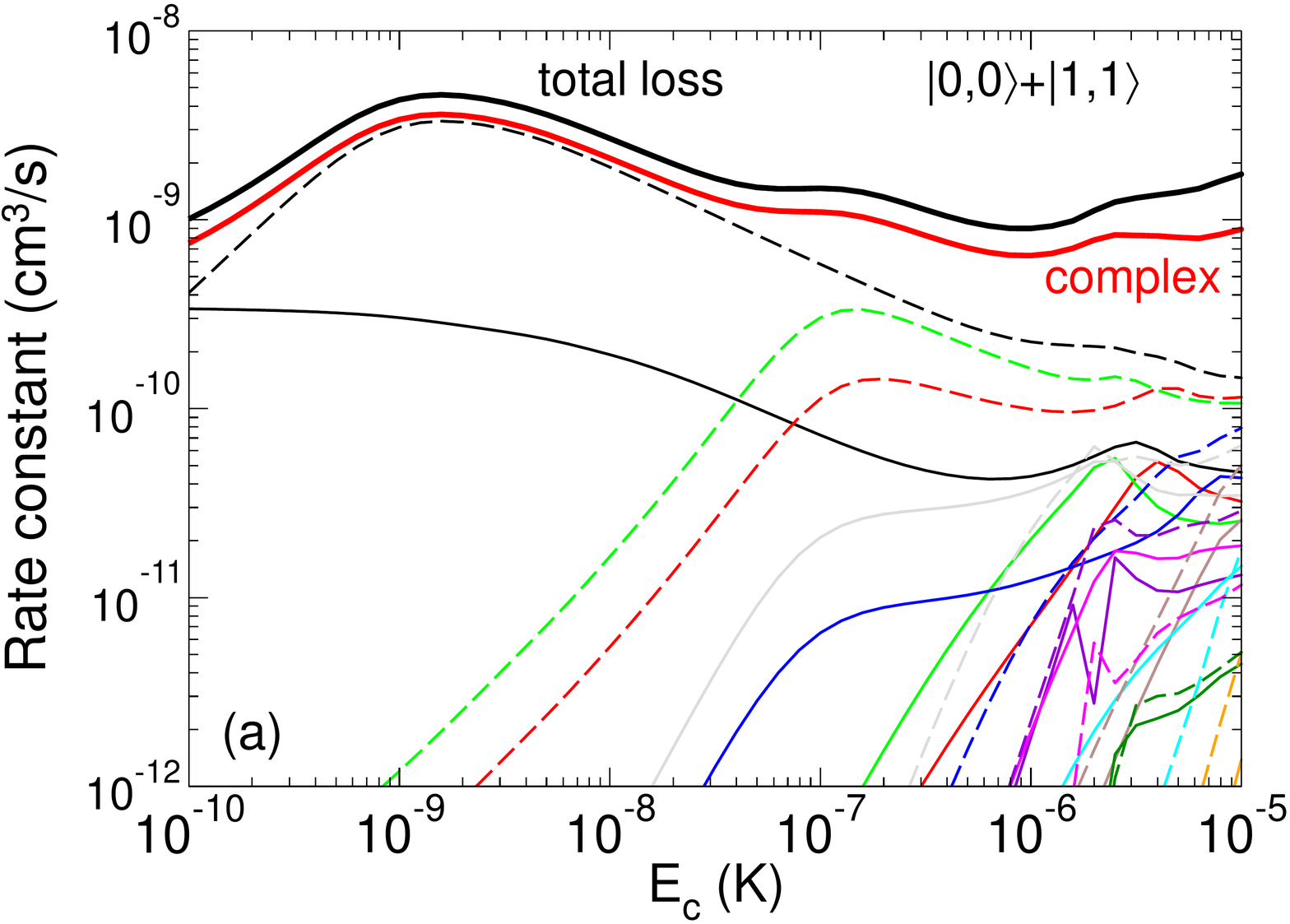} \\
\includegraphics[width = 0.9 \linewidth]{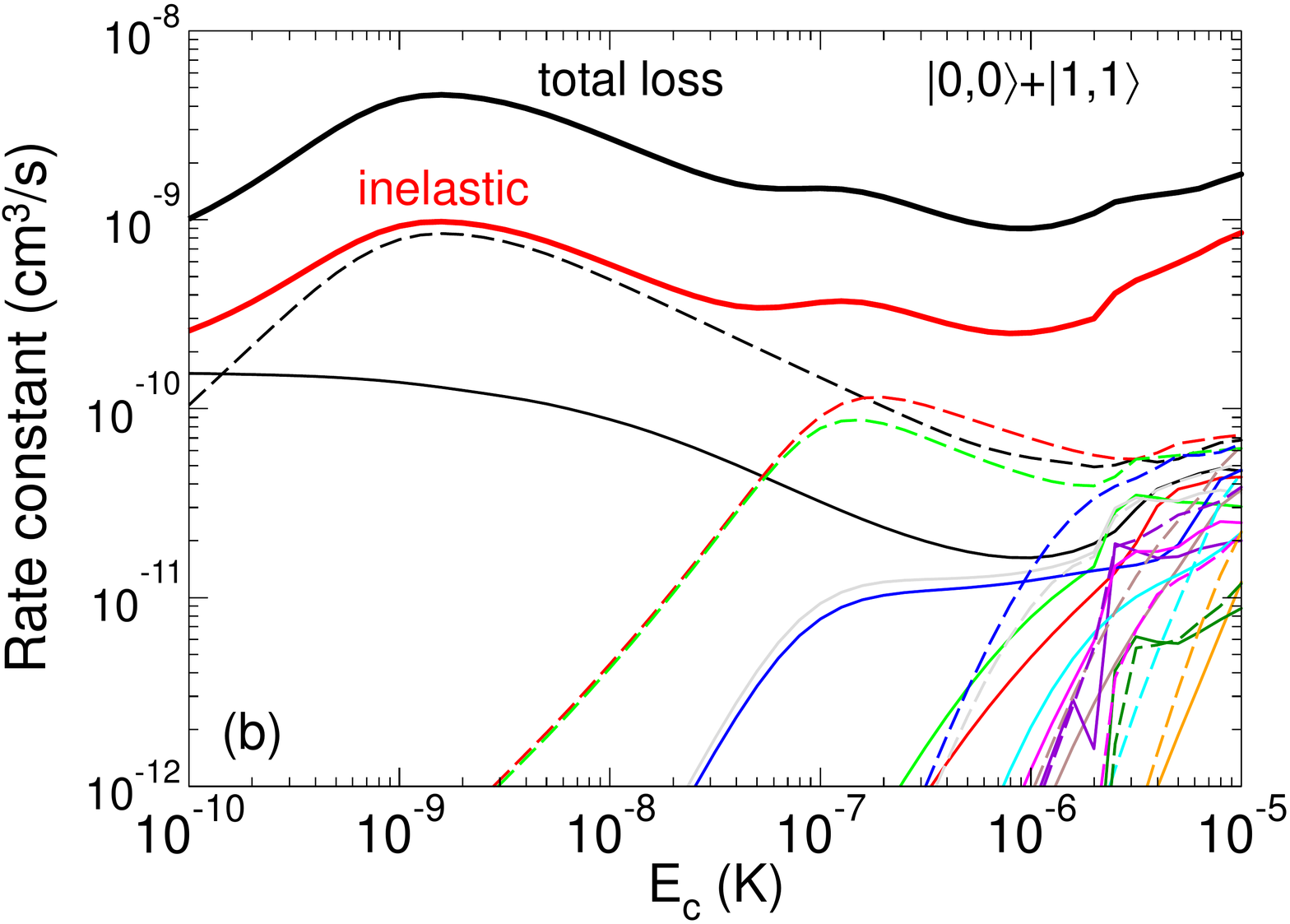}
\caption{
Loss rate constants $\beta_{01}$ for the mixture
$\ket{0,0}$ + $\ket{1,1}$ as a function of the collision energy. 
(a) corresponds to the complex formation loss rate constant
while (b) to the inelastic one.
Thin solid curves are for the symmetry $(\epsilon_i=-1, \eta=+1)$ corresponding to $l = 0,~2,... 10$. Thin dashed curves are for $(\epsilon_i=+1, \eta=-1)$ corresponding to $l = 1,~3,... 11$. 
Contributions from partial waves $m_l =-5,-4,...+5$ are included for both symmetries. 
The color code is the following: black curve ($m_l=0$), 
red curves ($m_l=+1$), green curves ($m_l=-1$),
blue curves ($m_l=+2$), grey curves ($m_l=-2$),
brown curves ($m_l=+3$), violet curves ($m_l=-3$),
cyan curves ($m_l=+4$), magenta curves ($m_l=-4$),
orange curves ($m_l=+5$), dark green curves ($m_l=-5$).
The red thick solid line in the (a) 
is the complex loss rate constant (sum of 22 curves).
In (b), this is the inelastic loss rate constant (sum of 22 curves).
The black thick solid line is the total loss rate constant and corresponds to the sum of all these 44 curves.}
\label{fig5}
\end{figure} 
 
For the $\ket{0,0}$ + $\ket{1,1}$ mixture, the $m_l =-5,-4,...+5$ values 
correspond to $M=2,3,...,11,12$ as $m_{F_1} = 3$ and $m_{F_2} = 4$.
The $m_l=0$ contribution corresponds to $M=7$. 
This results in 22 curves 
which are added to obtain the complex formation rate constant
[Fig.~\ref{fig5}(a)]
and 22 other curves 
which are added to obtain the inelastic, hyperfine state changing process rate constant
[Fig.~\ref{fig5}(b)].
The complex and inelastic rate constants are then added 
to obtain the total loss rate constant, reported in Fig.~\ref{fig5} 
as a function of the collision energy
and in Fig.~\ref{fig3}(c) as a function of the temperature.
The inelastic contribution is smaller than the complex formation rate at low collision energies
but becomes comparable at higher collision energies, typically around 10~$\mu$K. 
The increase in the inelastic rate between 2~$\mu$K and 10~$\mu$K is due to the opening
of an additional CMS corresponding to other hyperfine dressed states, as can be seen 
in the state-to-state rate constant next.
 
\subsection{State-to-state rate constants} 

\begin{figure}[h]
\centering
\includegraphics[width = 0.9 \linewidth]{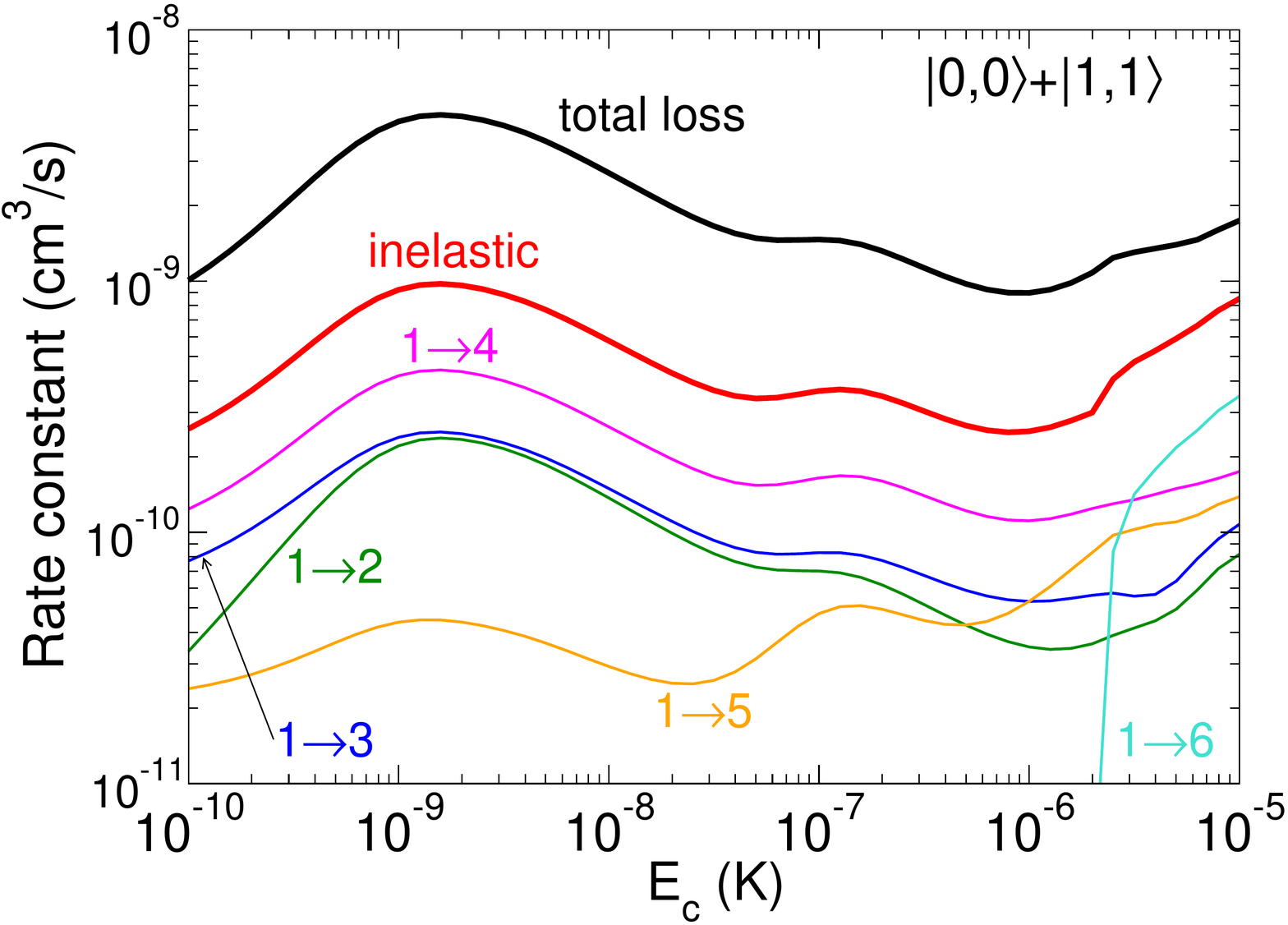} 
\caption{
State-to-state inelastic rate constants for the mixture
$\ket{0,0}$ + $\ket{1,1}$ as a function of the collision energy. 
The CMS notations are given in Eqs.~\eqref{CMSSTS}.
}
\label{fig6}
\end{figure}

For the $\ket{0,0}$ + $\ket{1,1}$ mixture, we also provide the contribution of the
different inelastic state-to-state rate constants of the total inelastic rate.
This is displayed in Fig.~\eqref{fig6} for the initial CMS noted 1 to all final states.
The list of CMS corresponds to
\begin{eqnarray}
1 &=& \{|0, 0, 3/2, 3/2 \rangle \, |1, 1, 3/2, 3/2\rangle \} \nonumber \\
2 &=& \{|0, 0, 3/2, 3/2 \rangle \, |1, 0, 3/2, 3/2\rangle \} \nonumber \\
3 &=& \{|0, 0, 3/2, 3/2 \rangle \, |1, -1, 3/2, 3/2\rangle \} \nonumber \\
4 &=& \{|0, 0, 3/2, 1/2 \rangle \, |1, 0, 3/2, 3/2\rangle \} \nonumber \\
5 &=& \{|0, 0, 3/2, 3/2 \rangle \, |1, -1, 3/2, 1/2\rangle \} \nonumber \\
6 &=& \{|0, 0, 3/2, 1/2 \rangle \, |1, -1, 3/2, 3/2\rangle \} .
\label{CMSSTS}
\end{eqnarray}
At collision energies below 2~$\mu$K, 
the inelastic processes correspond to relaxation ones to the
energetically lower open CMS 1 to 5,
while at collision energies above, as energetically higher CMS open up such as 6, 
inelastic excitation processes are also possible, explaining the increase
in the inelastic rate. 

\subsection{Comparison with experiments}

To compare with the experimental results, the theoretical total loss rate constant for the $\ket{0,0}$ + $\ket{1,0}$ mixture as a function of the temperature is presented in Fig.~\ref{fig3}(c) as a black dashed line. For the $\ket{0,0}$ + $\ket{1,1}$ mixture, this is presented as a red dashed line. An overall good agreement with the experimental results is found, both in magnitude and temperature dependence, considered the number of curves computed and added at the end, and the absence of any adjusting parameters. For the $\ket{0,0}$ + $\ket{1,1}$ mixture, the contribution of the complex formation loss only is presented as an orange dash-dot line and that from inelastic, hyperfine changing is shown by the orange short-dash curve. We can see that taking only into account the complex formation loss is not sufficient to get a proper agreement with the experimental data.  While the inelastic contribution is 2 to 3 times smaller than the complex loss, it is necessary to account for it
to explain the magnitude of the experimental rates. 
\\

\begin{figure}[h]
\centering
\includegraphics[width=1.0 \linewidth]{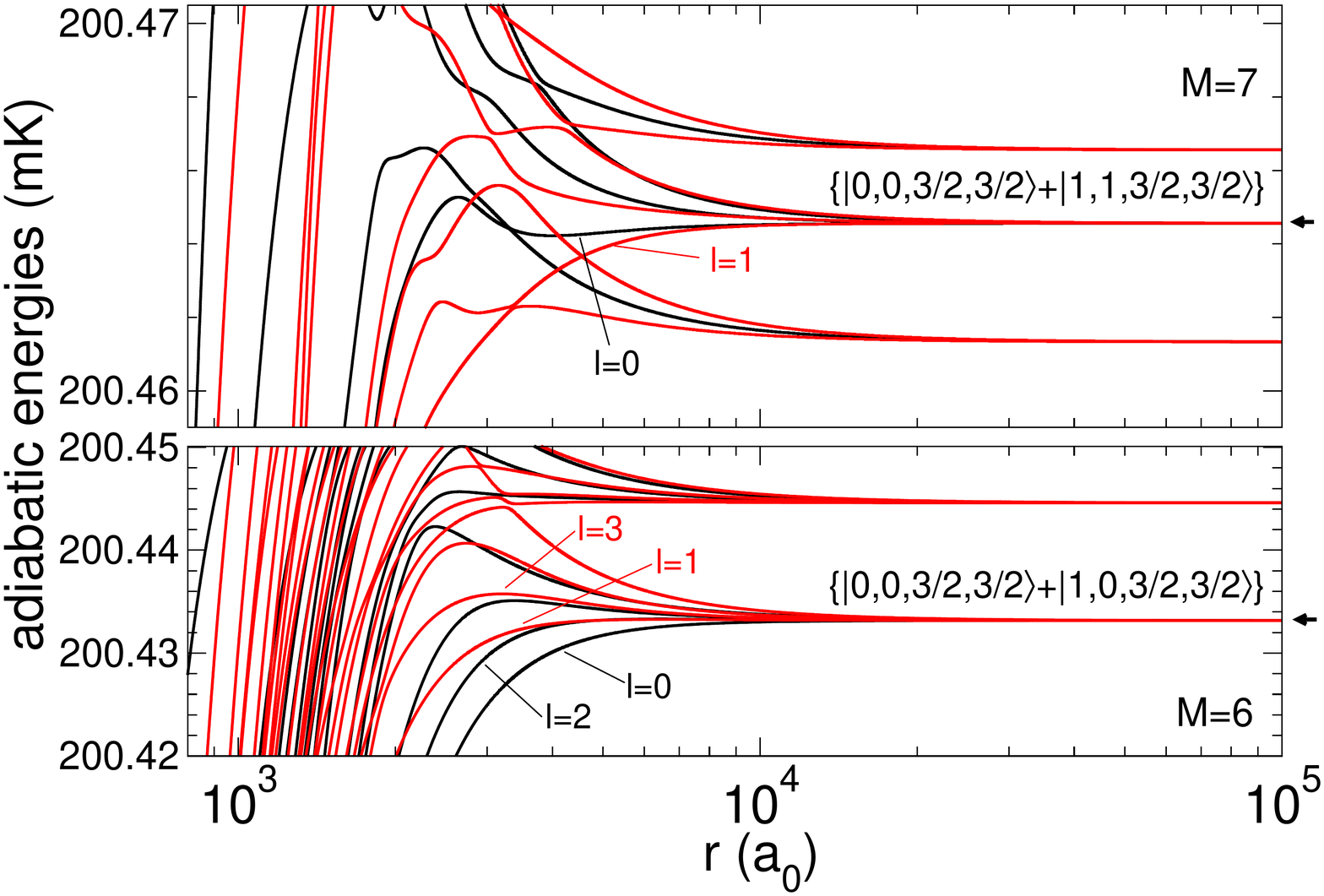}
\caption{Adiabatic energies for the $m_l=0$ scattering channels of the $\ket{0,0}$ + $\ket{1,1}$ (top panel) and $\ket{0,0}$ + $\ket{1,0}$ (bottom panel) mixtures. Each case corresponds respectively to a total projection quantum number $M =7$ and $M=6$. The arrow indicates the entrance channels. Even values of $l$ (black lines) and odd values of $l$ (red lines) belong to different symmetries, respectively the $(\epsilon_i=-1, \eta=+1)$ and $(\epsilon_i=+1, \eta=-1)$ symmetry.}
\label{fig7}
\end{figure}

The difference in the rate constant magnitudes between the two mixtures can be explained by the quantum calculation, as it takes into account the proper
specific, complicated dressing of the hyperfine eigenstates in the magnetic field and the strong interplay of the collisional channels via the DDI. This can be seen in Fig.~\ref{fig7} where the adiabatic energies, which act as effective potentials for the entrance collisional channels indicated by arrows, are presented for the component $m_l=0$.
For the $\ket{0,0}$ + $\ket{1,0}$ mixture, the entrance scattering channel is the lowest one. Therefore it is only affected by other scattering channels above. 
The channel is mainly governed by an attractive DDI and repulsive centrifugal terms, giving rise to the usual centrifugal barriers as the partial wave quantum number $l$ increases. This can be clearly seen in the bottom panel of Fig.~\ref{fig7} for $l=0,2$ and for $l=1,3$. 
The barrierless, attractive $s$-wave curve ($l=0$) provides the main and largest contribution to the rate constant, as it can be seen in Fig.~\ref{fig4} where the thin solid black curve $m_l=0, l=0$ dominates.
\\

In contrast, for the $\ket{0,0}$ + $\ket{1,1}$ mixture, the entrance scattering channel is higher in energy and lies between many other scattering channels of different hyperfine eigenstates. Because of the strong interplay between the DDI and the hyperfine couplings, the surrounding channels modify the entrance channel of this mixture significantly. As can be seen in the top panel of Fig.~\ref{fig7}, the $s$-wave curve becomes somewhat pushed up from the channels below, while the $p$-wave curve ($l=1$) becomes somewhat pushed down from the channels above and more attractive. Therefore, the $s$-wave curve contribute much less in this case and the $p$-wave curve contributes the most instead. 
This can also be seen in Fig.~\ref{fig5} where the thin dashed black curve for $m_l=0, l=1$ dominates
over the thin solid black curve for $m_l=0, l=0$, for both the complex loss process and the inelastic one, in the energy range corresponding to the experimental temperatures.
The overall rate constant is smaller for this mixture as the $p$-wave attractive curve is still less attractive than the $s$-wave one for the other mixture. This shows that, apart from the lowest scattering channel, collisions are highly specific to the chosen hyperfine eigenstates of the mixture, i.e., it is possible to further control the resonant DDI by selecting different hyperfine states.


\section{Conclusion}
We observed the resonant dipole-dipole collisions in incoherent rotational mixtures of ultracold \NaRb molecules. We also presented in full detail a quantum close-coupling model using a pertinent scheme of selected states to keep the numerical calculation feasible, which explains the experimental results very well. This joint experimental and theoretical investigation leads to a full understanding on the strongly dipolar molecular collisions mediated by complex formation. It also provides valuable insights into proposed applications with UPMs by harnessing the resonant DDI, such as using them as qubits for quantum information processing~\cite{ni18,Sawant2020,Hughes2020}. For many of these applications, the multi-channel nature, both internally and externally, of the strongly DDI between UPMs should be carefully taken into account. For instance, for single molecules not in the lowest lying rotational pairs and trapped in optical tweezers, while the probability for hyperfine changing to occur is low when the separation between the tweezers is large, it may become a problem if the molecules were brought to a small separation to enhance the DDI strength for faster gate operation~\cite{Caldwell2020}. This may introduce unwanted leakage which will reduce the fidelity of quantum gates. A possible remedy for this issue is by applying a small dc electric field to decouple the hyperfine levels and thus suppress the possible leakage channels. This idea has been successfully adopted to extend the coherent time in microwave coupled rotational levels~\cite{luo2018,blackmore2020}. 

Together with our previous work carried out in electric fields~\cite{guo18}, the current work shows that the universal model with unity short-range complex loss probability can better explain the dipolar UPM collisions than the non-dipolar UPM collisions~\cite{ye18,gre19,Bai2019}. This indicates that the UPM collisions become more universal in presence of the strong dipolar interaction. This may be understood by the enhanced complex density of state which extends the complex lifetime and increases the probability of complex loss~\cite{chris19,chris19a}.

\section*{Acknowledgments}
This work was supported by the Hong Kong RGC General Research Fund (grants 14301119, 14301818 and 14301815) and the Collaborative Research Fund C6026-16W. 
G.Q. acknowledges funding from the FEW2MANY-SHIELD Project No.~ANR-17-CE30-0015 and the COPOMOL Project No.~ANR-13-IS04-0004 from Agence Nationale de la Recherche. 

\section*{Appendix}

\subsection{Admixture of different hyperfine dressed states of interest}

We provide here a list in increasing energies of different hyperfine dressed states of interest 
at a magnetic field of $B=335$~G.
The dressed states are a linear combination of the bare states $\ket{J, m_J, m_I^{\rm Na}, m_I^{\rm Rb}}$ with different amplitudes (admixtures). 

\begin{widetext}
\begin{eqnarray} \label{hypdressstates}
\begin{split}
\{{\ket{0,0,3/2,3/2}}\}  &= - \ket{0,0,3/2,3/2}  \\ 
\{| 0, 0, 3/2, 1/2\rangle\} &=   -0.994 \, | 0, 0, 3/2, 1/2\rangle - 0.11 \, | 0, 0, 1/2, 3/2 \rangle \\
\{{\ket{1,0,3/2,3/2}}\}  &= - 0.906 \ket{1,0,3/2,3/2} + 0.422 \ket{1,1,3/2,1/2} \\
& +  0.011 \ket{1,1,1/2,3/2} \\
\{| 1, -1, 3/2, 3/2\rangle\} &=  -0.841 \, | 1, -1, 3/2, 3/2\rangle \\
& + 0.435 \, | 1, 1, 3/2, -1/2\rangle  - 0.313 \, | 1, 0, 3/2, 1/2\rangle \\
& - 0.064 , | 1, 0, 1/2, 3/2\rangle + 0.027 \, | 1, 0, 1/2, 1/2\rangle \\
\{| 1, 0, 1/2, 3/2\rangle\} &=  - 0.898 \, | 1, 0, 1/2, 3/2\rangle + 0.434 \, | 1, 1, 1/2, 1/2\rangle \\ 
& + 0.053\, | 1, -1, 3/2, 3/2\rangle  - 0.037 \, | 1, 1, 3/2, -1/2\rangle \\
& + 0.025\, | 1, 0, 3/2, 1/2\rangle  \\
\{{\ket{1,1,3/2,3/2}}\}  &= \ket{1,1,3/2,3/2} .
\end{split} \nonumber
\end{eqnarray}
\end{widetext}

\subsection{Adiabatic energies for the full and selected nuclear spin state structure}


\begin{figure}[t]
\centering
\includegraphics[width=1.0 \linewidth]{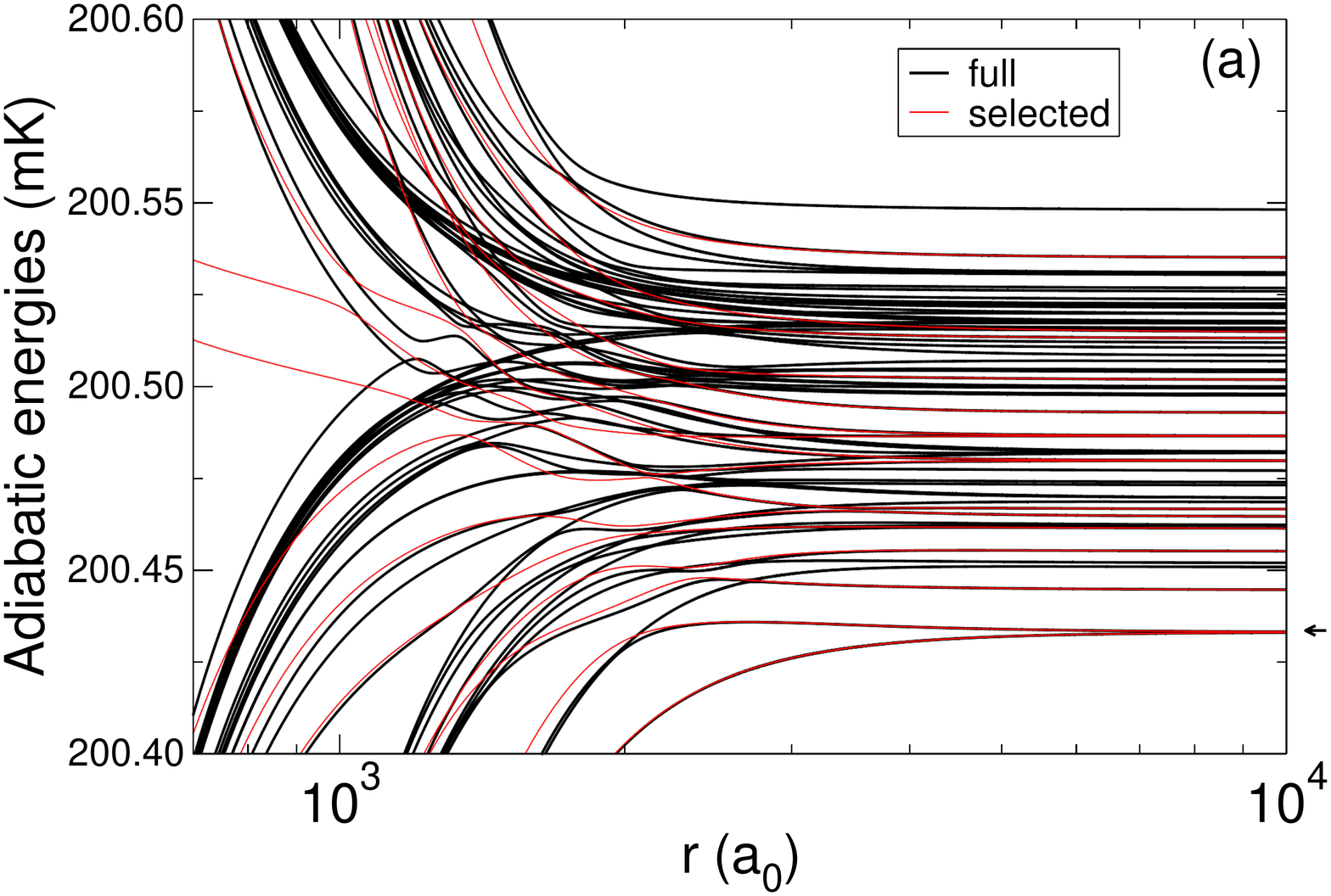} 
\includegraphics[width=1.0 \linewidth]{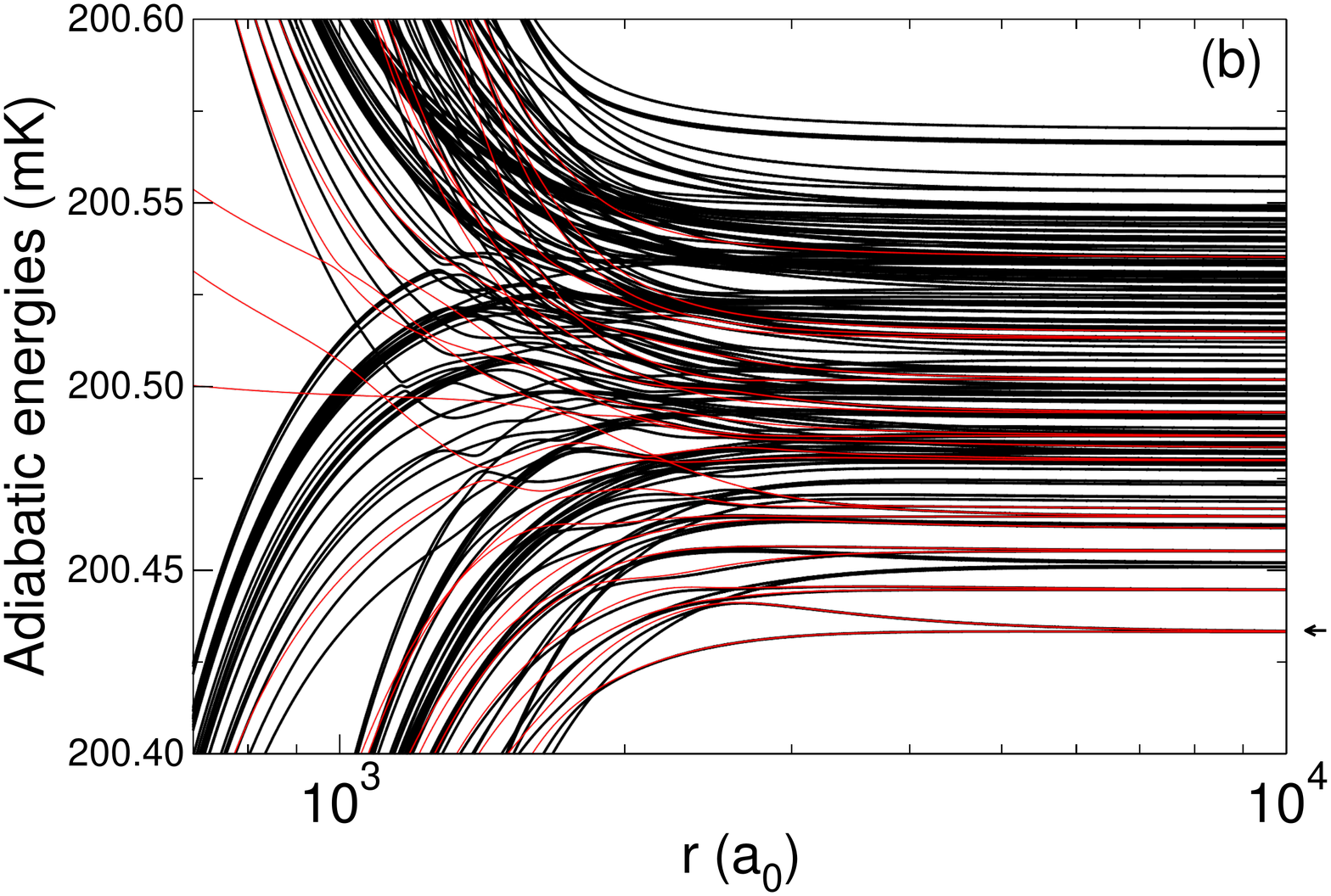}
\caption{Adiabatic energies for the $M=6$ projection for the symmetry ($\varepsilon_i=-1, \eta=+1$) in (a), and for the symmetry ($\varepsilon_i=+1, \eta=-1$) in (b).
The arrow indicates the initial CMS $\{|0, 0, 3/2, 3/2 \rangle \, |1, 0, 3/2, 3/2\rangle \} $ corresponding to the $\ket{0,0}$ + $\ket{1,0}$ mixture in this study. The $M=6$ projection corresponds to $m_l=0$ for this CMS.
The black curves calculation involves the full structure of hyperfine dressed states in the composition of the CMS while the red curves calculation involves relevant selected states.
Only two partial waves have been included to make the full structure calculation 
possible: $l=0$ and $l=2$ in (a) and $l=1$ and $l=3$ in (b).
}
\label{fig8}
\end{figure}

\begin{figure}[t]
\centering
\includegraphics*[width=1.0 \linewidth]{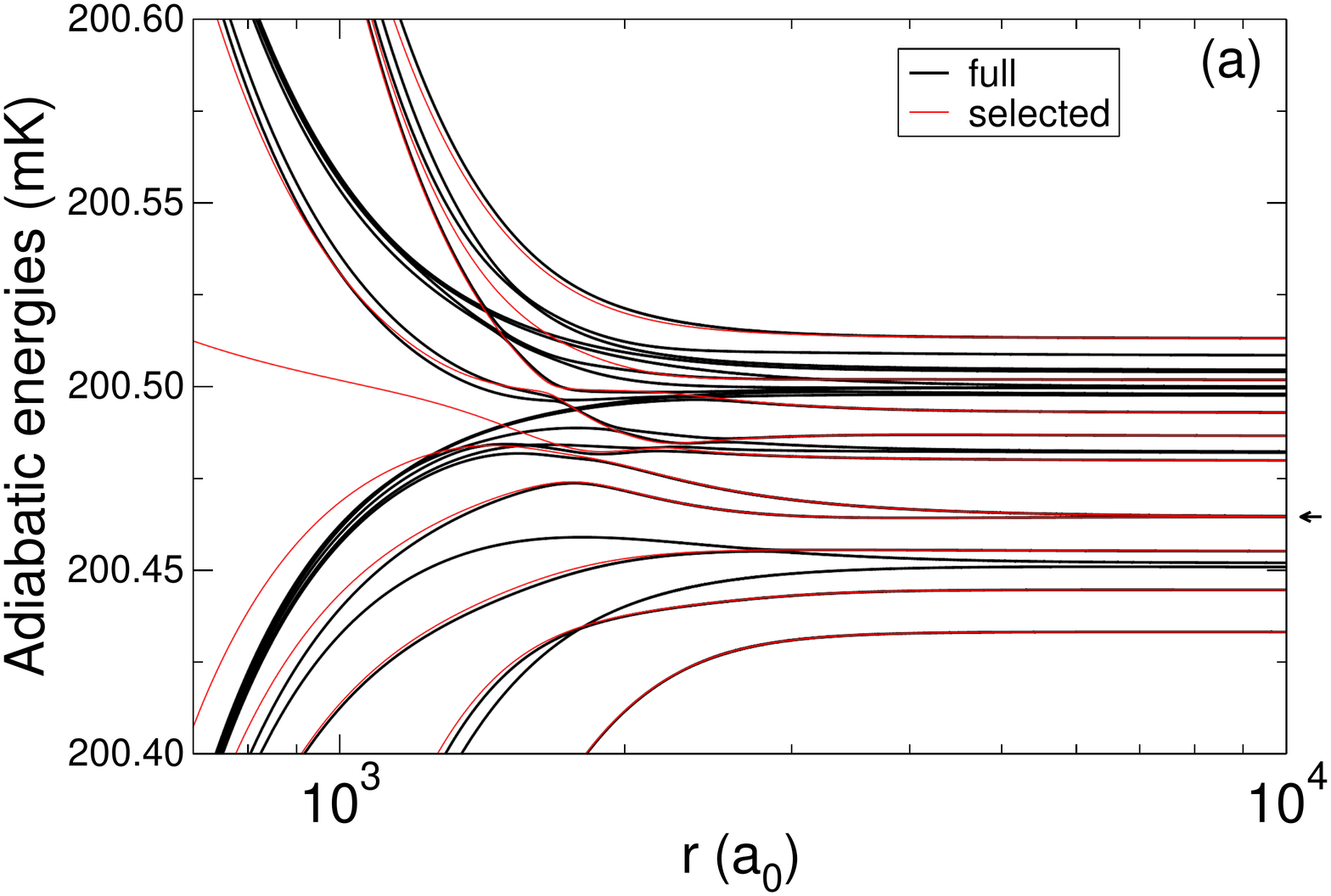}
\includegraphics*[width=1.0 \linewidth]{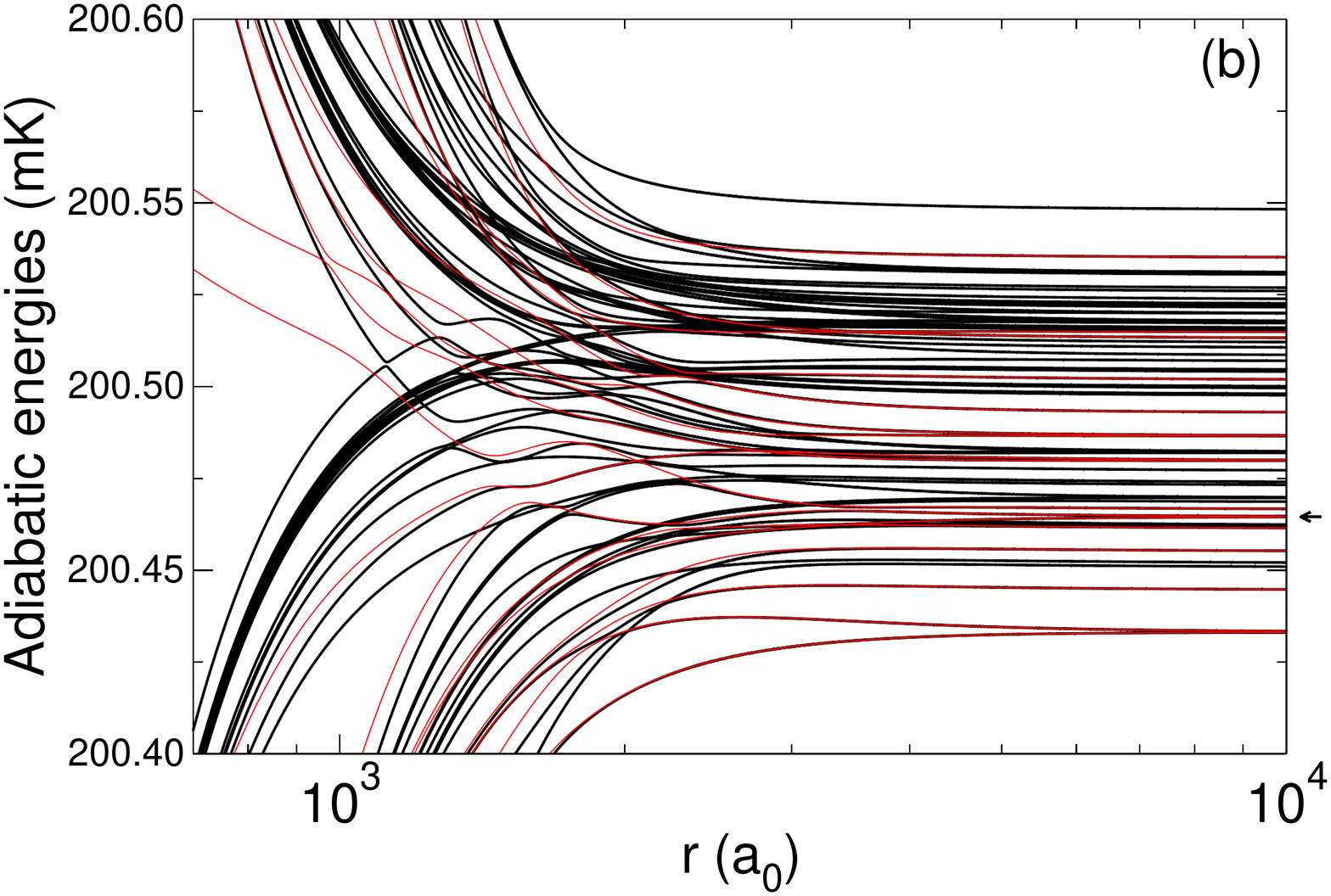}
\caption{Adiabatic energies for the $M=7$ projection for the symmetry ($\varepsilon_i=-1, \eta=+1$) in (a), and for the symmetry ($\varepsilon_i=+1, \eta=-1$) in (b).
The arrow indicates the initial CMS $\{|0, 0, 3/2, 3/2 \rangle \, |1, 1, 3/2, 3/2\rangle \} $ corresponding to the $\ket{0,0}$ + $\ket{1,1}$ mixture in this study. The $M=7$ projection also corresponds to $m_l=0$ for this CMS.}
\label{fig9}
\end{figure}


We present the adiabatic energies as a function of the relative distance$r$
 between the two molecules in Fig.~\ref{fig8} for the $M=6$ projection, which 
for the CMS $\{|0, 0, 3/2, 3/2 \rangle \, |1, 0, 3/2, 3/2\rangle \}$
($\ket{0,0}$ + $\ket{1,0}$ mixture) corresponds
to the $m_l=0$ component. The energy of this initial CMS is indicated by an arrow.
Both symmetries ($\varepsilon_i=-1, \eta=+1$) 
and ($\varepsilon_i=+1, \eta=-1$) are included, in Fig.~\ref{fig8}(a) and (b) respectively,
for the two lowest partial waves, $l=0$ and 2 and for $l=1$ and 3 respectively.
The black curves consider no simplification in the list of hyperfine dressed states, that is no removal of unnecessary dressed states. We therefore consider all the possible dressed combined molecular states
for the given partial waves. The red curves correspond to the simplification which consists in 
relevant selected hyperfine dressed states. 
We see then that there are fewer CMS involved. Yet, the incoming curves of the initial colliding state (see the arrow) are still very well represented and present no substantial differences with the black curves when all hyperfine dressed states are included. For ultracold collision energies, the dynamics will therefore be barely affected when doing the selected states simplification, and reducing by a lot the numerical effort.
Similar conclusions hold for the adiabatic energies in Fig.~\ref{fig9} 
for the $M=7$ projection, which for the CMS $\{|0, 0, 3/2, 3/2 \rangle \, |1, 1, 3/2, 3/2\rangle \}$ 
($\ket{0,0}$ + $\ket{1,1}$ mixture) corresponds to the $m_l=0$ component. 
%


%

\end{document}